**Determining the Curie temperature of $La_{0.67}Sr_{0.33}MnO_3$ thin films**

By *H. Boschker, J. Verbeeck, R. Egoavil, S. Bals, G. van Tendeloo, M. Huijben, E.P. Houwman, G. Koster, D.H.A. Blank,* and *G. Rijnders\**

[*]     Dr. H. Boschker, Dr. M. Huijben, Dr. E.P. Houwman, Dr. G. Koster, Prof. D.H.A. Blank, Prof. G. Rijnders
        Faculty of Science and Technology and MESA+ Institute for Nanotechnology
        University of Twente
        7500 AE, Enschede, The Netherlands
        E-mail: a.j.h.m.rijnders@utwente.nl
        Dr. J. Verbeeck, R. Egoavil, Dr. S. Bals, Prof. G. van Tendeloo
        Electron Microscopy for Materials Science (EMAT)
        University of Antwerp
        2020 Antwerp, Belgium

**Determination of the Curie temperature**

When comparing a set of $La_{0.67}Sr_{0.33}MnO_3$ (LSMO) samples, the Curie temperature ($T_C$) of the samples is an important figure of merit for the sample quality. Therefore, a reliable method to determine $T_C$ is required. Here, a method based on the analysis of the magnetization loops is proposed.

**Magnetization measurements**

$T_C$ is the temperature above, which the spontaneous magnetization disappears. However, the magnetic signal of LSMO does not drop to zero at $T_C$ in most measurements, as typically a small background field is used. This background field is necessary as LSMO's magnetization is very weak with a coercivity, which approaches zero around $T_C$. LSMO's spontaneous magnetization will therefore be divided into domains with different orientations and no net magnetization can be observed without the background field. Above $T_C$, LSMO has a very large magnetic susceptibility, as it is possible to induce ferromagnetic exchange coupling with the field. Therefore, a magnetization curve will typically show a tail of magnetic signal above $T_C$.

As a result, the precise transition temperature is open to interpretation. In literature, it is generally not mentioned which criterium is used. E.g., Kourkoutis *et al.* presented a magnetization curve from which a $T_C$ of approximately 298 K was derived.[1] Inspection of the curve, however, shows that 298 K is the temperature at which the tail of the magnetization vanishes. Alternatively, Molegraaf *et al.* presented magnetization curves where the $T_C$ was determined using a linear fit to the slope of the magnetization curve below $T_C$.[2] The intercept of the line with the $T$ axis was defined as $T_C$. We used a method based on the analysis of the temperature dependence of the magnetization loops.

In Fig. S1a and S1b, magnetization loops of the 8 unit cell layer LSMO sample for different temperatures close to $T_C$ are presented. All curves show a magnetic saturation with a reduction of the magnetic signal around remanence. A finite coercivity and remanent magnetization are not observed. This is either due to the absence of ferromagnetism (above $T_C$) or to domain formation (below $T_C$). In order to determine $T_C$, the low field part of the magnetization loops, as shown in Fig. S1b, was analyzed. At zero applied magnetic field the magnetization curves at 250 and 260 K increase rapidly. Then between 10 and 20 kA/m the curves increase less rapidly and linearly. The rapid increase is attributed to the alignment of domains with different orientations, while the linear increase is attributed to magnetization rotation away from the easy axis to the direction of measurement. The linear part of the magnetization curve can be extrapolated to zero applied field, as shown in Fig. S1b. The intercept of the line with the magnetization axis signifies the remanence the sample would have without the domain formation at low applied field, $M_{Rem, extr}$. With the definition of $T_C$ as the temperature at which the spontaneous magnetization, and therefore the remanent magnetic signal, disappears, $M_{Rem, extr}$ can be used for the determination of $T_C$. The criterium for $T_C$ used here is the lowest temperature at which $M_{Rem, extr} = 0$. For the sample shown in Fig. S1, $M_{Rem, extr}$ is almost zero at 270 K and $T_C$ is determined at 270 K.

Fig. S1c shows the saturation and $M_{Rem, extr}$ of the sample. Above $T_C$, the saturation is zero, while the magnetic signal observed in the measurements is indicated as well. The signal above $T_C$ persists at least up to 310 K. The expected temperature dependence of a Weiss ferromagnet[3,4], is shown as well. Three curves are drawn with a $T_C$ of 260, 270 and 280 K respectively. Only the curve with a $T_C$ of 270 K fits the data. A linear fit to the magnetization curve below $T_C$ is indicated as well. The intercept of this line with the $T$ axis is 285 K. So, the determination of $T_C$ with the method outlined here results in a lower value for $T_C$ with respect to the reports in literature.

**Transport measurements**

In Fig. S2, the transport measurements of the sample are presented. LSMO has a metal insulator transition at $T_C$ and this transition temperature can in principle be obtained from the temperature dependence of the resistivity. The first criterium which is used is the temperature at which the resistivity is maximum[5,6], which corresponds to 310 K for this sample. The second criterium is the temperature at which the derivative of the resistivity is maximum, 260 K for this sample. Finally, the magnetoresistance is expected to be maximal at $T_C$.[7] The magnetoresistance is shown in Fig. S2b and the maximum occurs at 270 K, which corresponds well with the $T_C$ determined from the magnetization curves. It was found that the maximum in magnetoresistance corresponded with the $T_C$ determined from the magnetization loops for all samples measured.

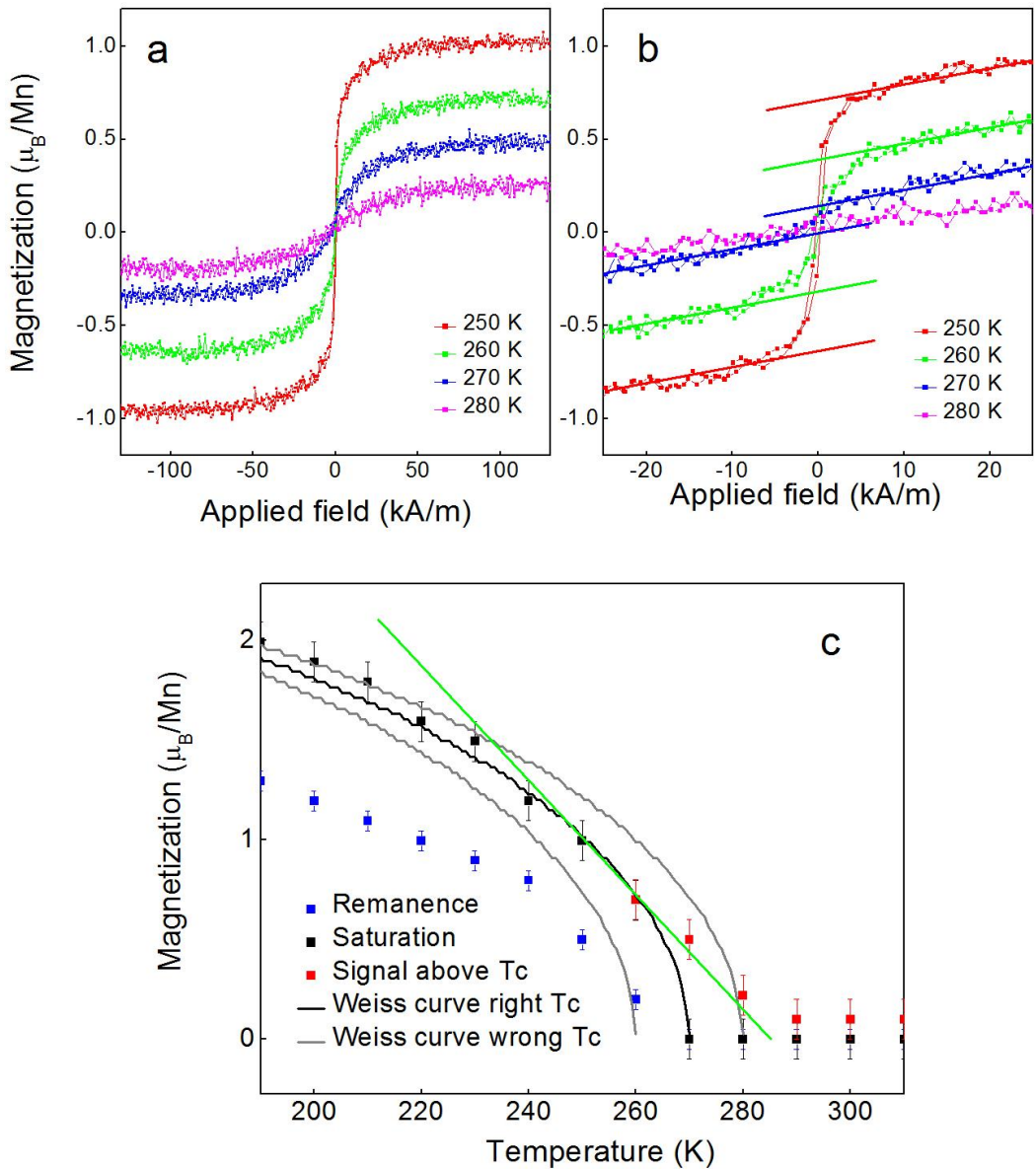

**Figure S1.** Magnetization measurements of an 8 unit cell layer LSMO film. The magnetic field was applied parallel to the $[100]_{pc}$ axis. a) Magnetization loops for different temperatures around $T_C$. b) same as a), but at low applied field. c) Temperature dependence of the magnetization. The remanence plotted is the extrapolated remanent signal, $M_{Rem, extr}$, as discussed in the main text. The green line is a linear extrapolation of the magnetization curve just below $T_C$.

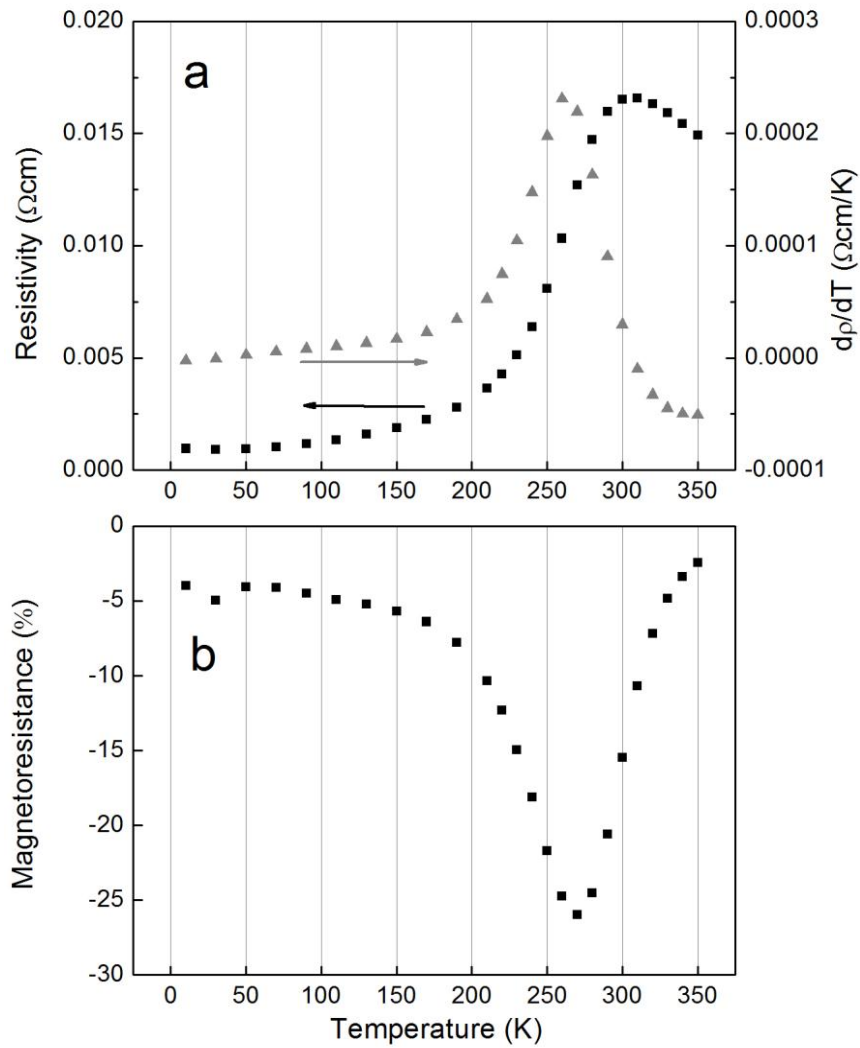

**Figure S2.** Transport measurements of an 8 unit cell layer LSMO film. a) Resistivity and its derivative. b) Magnetoresistance: ($\rho_{2.5T}$ - $\rho_{0T}$)/ $\rho_{0T}$.